**Title:**

Mass measurements of $^{99\text{-}101}$In challenge ab initio nuclear theory of the nuclide $^{100}$Sn.

**Author list:**

M. Mougeot,[1,2] D. Atanasov,[2] J. Karthein,[1,2] R. N. Wolf,[3] P. Ascher,[4] K. Blaum,[1]
K. Chrysalidis,[2] G. Hagen,[5,6] J.D. Holt,[7,8] W.J. Huang,[1] G.R. Jansen,[9] I. Kulikov,[10]
Yu. A. Litvinov,[10] D. Lunney,[11] V. Manea,[2,11] T. Miyagi,[7] T. Papenbrock,[5,6] L. Schweikhard,[12]
A. Schwenk,[13,14,1] T. Steinsberger,[1] S.R. Stroberg,[15] Z. H. Sun,[5,6] A. Welker,[2]
F. Wienholtz,[2,12,13] S.G. Wilkins,[2] and K. Zuber [16]

**Affiliations:**

1 Max-Planck-Institut für Kernphysik, Saupfercheckweg 1, 69117 Heidelberg, Germany
2 CERN, 1211 Geneva, Switzerland
3 ARC Centre of Excellence for Engineered Quantum Systems, School of Physics, The University of Sydney, NSW 2006, Australia
4 Centre d'Etudes Nucléaires de Bordeaux Gradignan, UMR 5797 CNRS/IN2P3 - Université de Bordeaux, 19 Chemin du Solarium, CS 10120, F-33175 Gradignan Cedex, France
5 Department of Physics and Astronomy, University of Tennessee, Knoxville, Tennessee 37996, USA
6 Physics Division, Oak Ridge National Laboratory, Oak Ridge, Tennessee 37831, USA
7 TRIUMF, 4004 Wesbrook Mall, Vancouver, BC V6T 2A3, Canada
8 Department of Physics, McGill University, 3600 Rue University, Montreal, QC H3A 2T8, Canada
9 National Center for Computational Sciences, Oak Ridge National Laboratory, Oak Ridge, Tennessee 37831, USA
10 GSI Helmholtzzentrum für Schwerionenforschung GmbH, 64291 Darmstadt, Germany
11 CNRS/IN2P3, IJCLab, Université Paris-Saclay, 91405 Orsay, France
12 Institut für Physik, Universität Greifswald, 17487 Greifswald, Germany
13 Technische Universität Darmstadt, Department of Physics, 64289 Darmstadt, Germany
14 ExtreMe Matter Institute EMMI, GSI Helmholtzzentrum für Schwerionenforschung GmbH, 64291 Darmstadt, Germany
15 Physics Department, University of Washington, Seattle, WA 98195, USA
16 Technische Universität Dresden, 01069 Dresden, Germany

**Abstract:**

**The tin isotope $^{100}$Sn is of singular interest for nuclear structure due to its closed-shell proton and neutron configurations. It is also the heaviest nucleus comprising protons and neutrons in equal number — a feature that enhances the contribution of the short-range proton–neutron pairing interaction and strongly influences its decay via the weak interaction. Decay studies in the region of $^{100}$Sn have attempted to prove its doubly magic character[1] but few have studied it from an *ab initio* theoretical perspective[2,3] and none of these has addressed the odd-proton neighbours, which are inherently more difficult to describe but crucial for a complete test of nuclear forces. Here we present direct mass measurement of the exotic odd-proton nuclide $^{100}$In, the beta-decay daughter of $^{100}$Sn, and of $^{99}$In with one proton less than $^{100}$Sn. We use advanced mass spectrometry techniques to measure $^{99}$In, which is produced at a rate of only a few ions per second, and to resolve the ground and isomeric states in $^{101}$In. The experimental results are compared with ab initio many-body calculations. The 100-fold improvement in precision of the $^{100}$In mass value highlights a discrepancy in the atomic mass values of $^{100}$Sn deduced from recent beta-decay results[4,5].**

The nuclear landscape is shaped by the underlying strong, weak, and electromagnetic forces. The most salient features are the pillars of enhanced differential binding energy associated with closed shell configurations, the most striking example of which is $Z = 50$ (tin), featuring the largest number of $\beta$-stable isotopes (10) of all elements. These nuclides lie between the closed neutron shells $N = 50$ and 82, conferring particular importance to the nuclides $^{100}$Sn and $^{132}$Sn. The neutron-rich $^{132}$Sn can be synthesized in comfortable quantities[6]. This is not so for $^{100}$Sn, forming the limit of proton stability due to its extreme neutron deficiency, only just staving off the Coulomb repulsion of the 50 protons. This rare combination of like closed shells causes $^{100}$Sn to have one of the strongest beta transitions and makes it the heaviest self-conjugate nucleus on the nuclear chart.

Nuclei in the immediate vicinity of $^{100}$Sn offer important insight for understanding the single neutron and proton states in this region and constitute an excellent proxy for the study of $^{100}$Sn itself. However, experiments were so far only feasible with in-beam gamma-ray spectroscopy performed at ~~fusion-evaporation and~~ fragmentation facilities[4,5,7–10]. By direct determination of the nuclear binding energy, high-precision atomic mass measurements provide a crucial model-independent probe of the structural evolution of exotic nuclei. Precision mass measurements are traditionally performed at Isotope Separation On Line (ISOL) facilities, however the production of medium-mass, neutron-deficient nuclides at such facilities is prohibitively difficult, explaining the lack of accurate mass values in the region. Measurements performed recently at the FRS-Ion Catcher at GSI[11] and the CSRe storage-ring in Langzhou[12] (both high-energy, heavy-ion fragmentation facilities) recently extended direct mass measurements to the $^{101}$In ground and isomeric states. However, the $^{100}$In mass value is still constrained 63% indirectly through its beta-decay link to $^{100}$Cd[13].

Thus, the first experimental challenge overcome in this work was the production and separation of the successfully studied $^{99,100,101g,101m}$In states. A detailed schematic of the necessary stages, from radioactive ion beam production to beam purification, preparation and measurement, is shown in Fig. 1. The exotic indium isotopes were produced at the ISOLDE on-line isotope separator located at CERN. A 1.4-GeV proton beam impinged on a thick lanthanum carbide target, producing a swath of neutron-deficient radioactive species of various chemical elements. After diffusion from the heated target, the indium atoms of interest were selectively ionized using a two-step resonance laser ionization scheme provided by the ISOLDE Resonant Ionization Laser Ion Source (RILIS)[14]. The ion beam was extracted from the source and accelerated to an energy of 40 keV. The mass number ($A = Z+N$) of interest was selected by using ISOLDE's High-Resolution dipole mass Separator (HRS) and delivered to the ISOLTRAP on-line mass spectrometer[15].

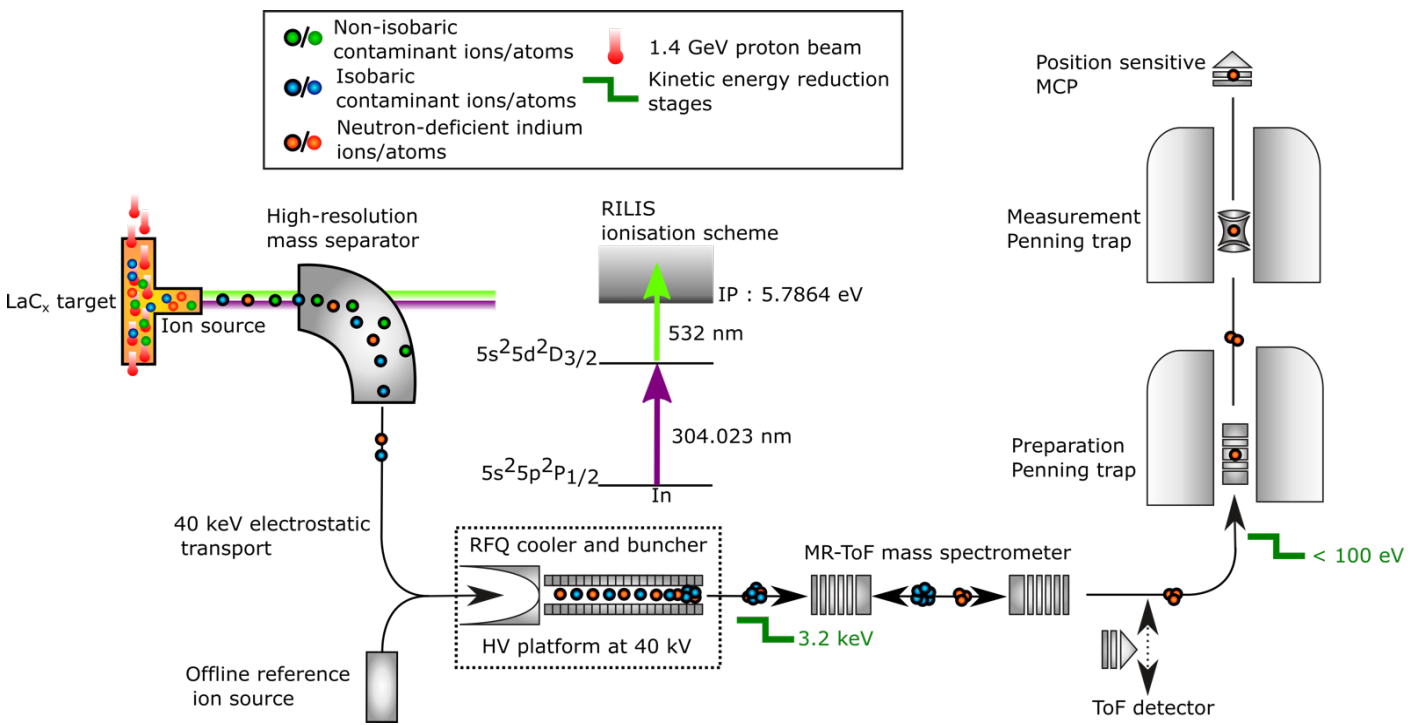

**Figure 1 | High-precision mass measurements of neutron-deficient indium isotopes with ISOLTRAP at**

**ISOLDE/CERN.** Radioactive atoms were produced by nuclear reactions of 1.4 GeV protons impinging on a thick lanthanum carbide target. Short-lived indium atoms diffusing from the target were selectively ionized using a two-step laser excitation scheme. The extracted ion beam was mass separated and injected into a radio-frequency ion trap, where it was bunched and cooled. The beam was then processed by a multi-reflection time-of-flight mass separator (MR-ToF MS) to separate the indium ions from the isobaric contaminants. When the precision Penning trap was used for the mass measurement, further cooling and purification of the beam was achieved using a helium buffer-gas-filled preparation Penning trap. In the case of $^{99}$In, for which the production yield was too low, the MR-ToF MS was used to perform the mass measurement. Reference alkali ions were provided by the ISOLTRAP offline ion source (see text for details).

The ions were first accumulated in ISOLTRAP's linear radio-frequency quadrupole cooler and buncher trap[16]. The extracted bunches were subsequently decelerated by a pulsed drift cavity to an energy of 3.2 keV before being purified by the multi-reflection time-of-flight mass separator (MR-ToF MS)[17], where multiple passages between two electrostatic mirrors rapidly separate the short-lived indium ions from much more abundant molecules of approximately the same mass. For all investigated isotopes, surviving molecular ions $^{80-82}$Sr$^{19}$F$^+$ were predominant in the ISOLDE beam. After a typical trapping time of about 25 ms, a resolving power in excess of m/Δm=10$^5$ was achieved. This combination of speed and high resolving power enables the MR-ToF to perform precise mass measurements of very short-lived species (see Methods). Because of its low production yield of < 10 ions s$^{-1}$, $^{99}$In was measured with this latter method only (see typical MR-ToF MS spectrum in Fig. 2).

The rate of $^{100}$In and $^{101}$In after the MR-ToF MS was sufficient to perform Penning-trap mass measurements. For $^{100}$In the conventional Time-of-Flight Ion-Cyclotron-Resonance (ToF-ICR) technique was used (see Methods). Even-$N$ neutron-deficient indium isotopes are known to exhibit long lived isomeric states lying a few hundred keV above the corresponding ground state, owing to the close energy proximity between the $\pi g_{9/2}$ and $\pi p_{1/2}$ states and their large spin difference. As a result, the $A = 101$ indium beam delivered to ISOLTRAP was a mixture of such two states so that the novel phase-imaging ion-cyclotron-resonance technique (PI-ICR)[18,19] had to be used to resolve them and ensure the accuracy of the ground-state mass value (see Methods for more details).

Table 1 summarizes our experimental results and compares them to the literature. The ISOLTRAP mass values for the ground and isomeric states of $^{101}$In agree well with averages obtained from Refs. [11,12]. The excitation energy is determined to be 668(11) keV, reducing the uncertainty by a factor of four. The ToF-ICR measurement of $^{101g}$In is in excellent agreement with the value measured using PI-ICR. $^{100}$In is found to be 130 keV more bound while the mass uncertainty is improved by almost a factor of 90. The mass of $^{99}$In is measured for the first time.

| | Half-life (s) | Method | Ref. nuclides | Ratio $r$ or $C_{ToF}$ | $M_E$ (keV) this work | $M_E$ (keV) Literature |
|---|---|---|---|---|---|---|
| $^{99}$In | 3.1(2) | MR-ToF MS | $^{80}$Sr$^{19}$F$^+$/$^{85}$Rb$^+$ | $C_{ToF}$ = 0.50076578(567) | -61429(77) | -61380#(300#) |
| $^{100}$In | 5.83(17) | MR-ToF MS | $^{81}$Sr$^{19}$F$^+$/$^{85}$Rb$^+$ | $C_{ToF}$ = 0.50060677(139) | -64187(20) | -64310(180) |
| | | ToF-ICR | $^{85}$Rb$^+$ | $r_{ref,x}$ = 1.1768824946(283) | -64178.2(22) | |
| $^{101g}$In | 15.1(11) | ToF-ICR | $^{85}$Rb$^+$ | $r_{ref,x}$ = 1.1886042835(590) | -68545.4(47) | -68545(12) |
| | | PI-ICR | $^{82}$Sr$^{19}$F$^+$ | $r_{ref,x}$ = 1.0000952633(432) | -68542.5(69) | |
| $^{101m}$In | 10# | PI-ICR | $^{82}$Sr$^{19}$F$^+$ | $r_{ref,x}$ = 1.0001023696(659) | -67874.5(83) | -67907(36) |

**Table 1 | Summary of the mass values obtained in this work.** (columns 1-7) isotope, half-life[20], measurement method (MR-ToF MS: multi-reflection time-of-flight mass spectrometer; ToF-ICR: time-of-flight ion-cyclotron resonance; or PI-ICR: phase-imaging ion-cyclotron resonance), reference (Ref.) nuclide used for the calibration, experimental frequency ratio $r$ or ToF constant $C_{ToF}$, and the resulting mass excess [$M_E = M$(atomic mass) – $A$(Atomic mass number) × $u$(atomic mass unit)]. For comparison, the results from the 2016 Atomic-Mass Evaluation (AME2016)[21] are listed for $^{99,100}$In (# indicates extrapolated mass value). For $^{101g,m}$In the values are the weighted average of two recent measurements performed at the FRS IonCatcher in GSI[11] and at the CSRe in Lanzhou[12]. The atomic mass values of the reference nuclides

are $m(^{85}Rb) = 84911789.738(5)$ µu, $m(^{81}Sr^{19}F) = 99921615 (3) $ µu, $m(^{82}Sr^{19}F) = 100916803(6)$ µu (from AME2016). The mass of the $^{80}Sr^{19}F$ reference was also measured during this run with the ToF-ICR technique using $^{85}Rb$ as reference yielding a frequency $r = 1.1650090659(365)$, as a result the corresponding $m(^{80}Sr^{19}F) = 98922914(3)$ µu was used.

Since the $^{100}Sn$ AME2016 mass excess value of -57280(300) keV[21] is derived from that of $^{100}In$ and the β-decay energy of Ref.[4], our new $^{100}In$ result improves the $^{100}Sn$ mass excess to -57148(240) keV. However, combining our result with the more recently published β-decay Q-value from Ref.[5] yields a $^{100}Sn$ mass excess of -56488(160) keV. For both decay energies, the $^{100}Sn$ mass is found to be more bound than previously inferred. In addition, the almost two standard deviations between the Q-values from Refs. [4,5] yield $^{100}Sn$ mass values which differ by 650 keV. We examine the consequences below and resolve this inconsistency.

Because the binding energy is a large quantity, finite differences are commonly used for assessing changes in nuclear structure from the mass surface. Shown in Fig. 2 (open grey symbols) is the two-neutron empirical shell-gap defined as $\Delta_{2n}(Z,N_0)=M_E(Z,N_0-2)-2M_E(Z,N_0)+M_E(Z,N_0+2)$, where $M_E(Z,N_0) = M_{atomic}(Z,N_0) – (Z+N_0) \times $ u(atomic mass unit) is the mass excess of a nucleus with $Z$ protons and a magic neutron number $N_0$. It shows a local maximum at the crossing of a magic proton number, a phenomenon, known as "mutually enhanced magicity"[22].

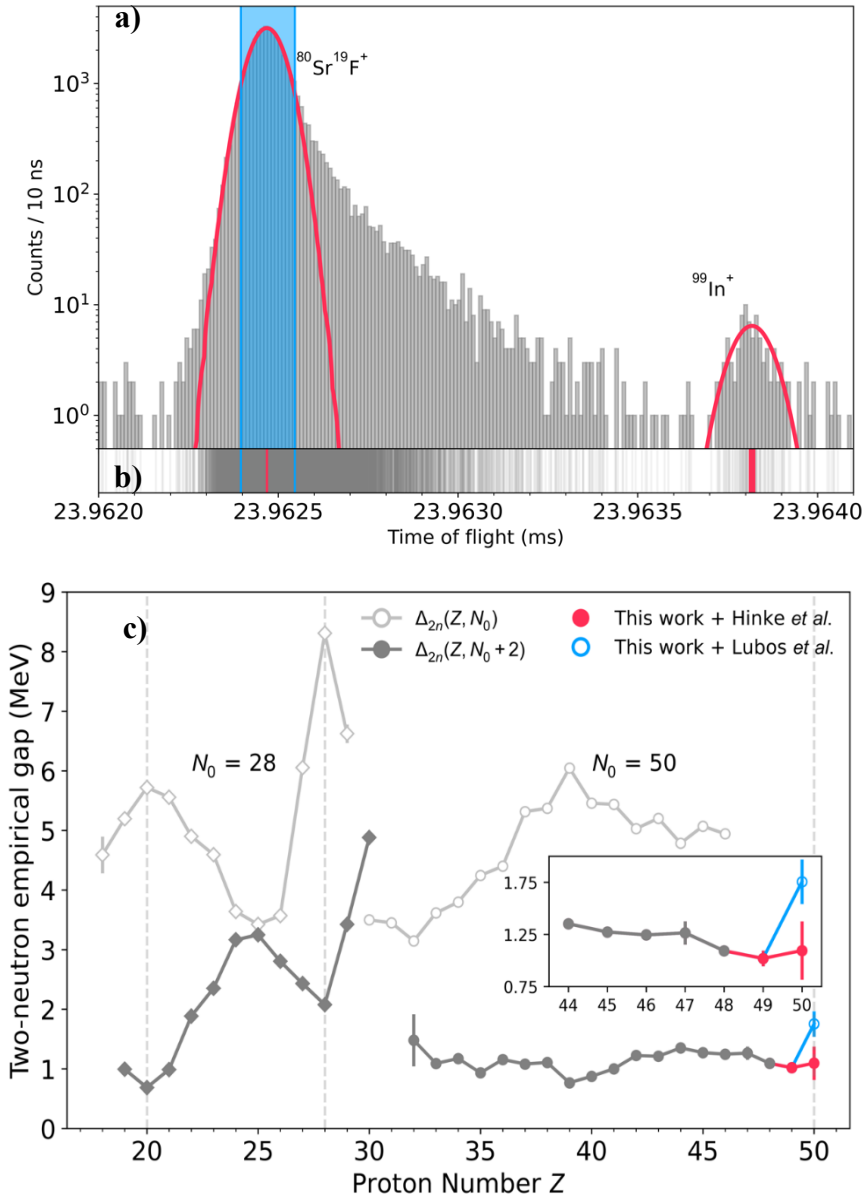

**Figure 2 | Overview of the experimental results**. **a)** A typical $A = 99$ MR-ToF MS spectrum obtained after 1000 revolutions. The solid red lines represent the Gaussian fit PDF scaled to the histogram within the used fit range. The blue band indicates the restricted fit range used for the $^{82}$Sr$^{19}$F peak analysis. **b)** Unbinned time-of-flight data used to perform the mass evaluation. The red vertical bars represent the uncertainty of the mean of the time-of-flight distributions at the ±1-sigma confidence level. An overview of the experimental data can be found in the Methods section. **c)** Empirical two-neutron shell gap $\Delta_{2n}(Z,N_0)$ as a function of proton number Z for the neutron magic numbers $N_0 = 28, 50$ (open grey symbols). The solid grey symbols show the corresponding value of the quantity $\Delta_{2n}(Z,N_0+2)$. At $Z = 50$ the solid red point corresponds to the value of $\Delta_{2n}(Z,N_0+2)$ calculated using the new masses from this work and the β-decay energy from Hinke et al.[4] while the open blue circle is using the value from Lubos et al.[5]. The inset shows a 2.5 magnification of the $\Delta_{2n}(Z,N_0+2)$ curve from $Z = 44$ to 50.

Since the lack of mass data for the $N=48$ isotopes of In ($Z=49$), Cd ($Z=48$) and Ag ($Z=47$) prevents deriving this quantity out to $^{100}$Sn we adapt an approach proposed in Ref. [23] using $\Delta_{2n}(Z,N_0+2)$, which is inversely correlated to $\Delta_{2n}(Z,N_0)$ (filled grey symbols in Fig. 2c). With this difference, a local *minimum* is observed because the binding energy of the magic neutron number appears in $\Delta_{2n}(Z,N_0+2)$ with opposite sign. The case of $N = 28$ is shown in Fig. 2 for illustration. Our new data allows extending $\Delta_{2n}(Z,N_0+2)$ to $Z = 49$ (indium) and indicates a slight downward trend towards $Z = 50$ (see inset in Fig. 2), as expected for a doubly magic $^{100}$Sn. Eliminating the contribution of the $^{100}$In ground-state mass uncertainty in the calculation of the $^{100}$Sn mass directly allows confronting the nuclear structure implications of the two Q-values from Refs.[4,5] and a global picture now emerges for this region. As shown, the Q-value reported by Lubos et al.[5] yields a $^{100}$Sn mass value which is at odds with the expected trend of $\Delta_{2n}(Z,N_0+2)$ to $Z = 49$ (open blue point in the bottom panel of Figure 2), whereas the value of Hinke et al.[4] yields a $^{100}$Sn mass which agrees with the trend within experimental uncertainties and is in line with our observation for $Z = 49$. In other words, while the Q-value reported in Ref. [4] follows the expectation of a doubly-magic $^{100}$Sn, the more recent (and higher statistics) Q-value reported in Ref. [5] yields a $^{100}$Sn mass value which suggests quite the opposite. Such a conclusion is at odds with new *ab initio* many-body calculations as discussed below.

In recent years, there has been great progress advancing *ab initio* calculations in medium-mass nuclei [24,25] up to the tin isotopes[2] based on modern nuclear forces derived from chiral effective field theory of the strong interaction, QCD. Most *ab initio* approaches are benchmarked on even-even nuclei, which are considerably simpler to compute, but this excludes from the benchmark effects that are only visible in odd nuclei. Among those are the single-particle states accessible to the unpaired nucleon and their interaction with the states of the even-even core, the blocking effect on pairing correlations and, in the case of odd-odd nuclei, the residual interaction between the unpaired proton and neutron. The latter two give rise to an odd-even-staggering (OES) of binding energies which can be quantified by a three-point estimator. Odd systems thus provide a complementary and stringent testing ground for novel theoretical approaches. Among *ab initio* approaches, the valence space formulation of the in-medium similarity renormalization (VS-IMSRG) group[26] is able to access a broad range of closed and open-shell nuclei in the nuclear chart[27]. In addition, we will explore the novel shell-model coupled-cluster method (SMCC)[28] for the first time in this region. Both the VS-IMSRG and coupled-cluster calculations provide access to a broad range of observables, such as first *ab initio* calculations of beta decays – up to $^{100}$Sn[3]. The VS-IMSRG was also recently shown to adequately describe both OES of nuclear masses and charge radii in neutron-rich odd-Z copper ($Z=29$) isotopes[29]. Here we present new VS-IMSRG and SMCC results that allow direct comparisons to the odd-Z nuclides adjacent to the iconic $^{100}$Sn nucleus.

We have performed cross-shell VS-IMSRG[30] and SMCC calculations using the 1.8/2.0 (EM) two-nucleon (NN) and three-nucleon (3N) interactions of Ref. [31]. This interaction is fit to the properties of nuclear systems with only $A = 2$, 3 and 4 nucleons (with 3N couplings adjusted to reproduce the triton binding energy and the $^4$He charge radius) and gives accurate results for ground-state energies of light and medium-mass nuclei[27,32]. To further explore the sensitivity to chiral EFT interactions, we also consider the NN + 3N(lnl) interaction[33] that has proven to constitute a valuable addition to existing chiral Hamiltonians in medium mass nuclei[34] but has yet to be tested in heavier systems. Finally, we show the first results for the $^{100}$Sn region with the ΔNNLO$_{GO}$(394) interaction[35]. Calculations with the ΔNNLO$_{GO}$(394) interaction and NN + 3N(lnl) were performed using the SMCC and VS-IMSRG methods, respectively. Technical details regarding these computations can be found in the Methods section.

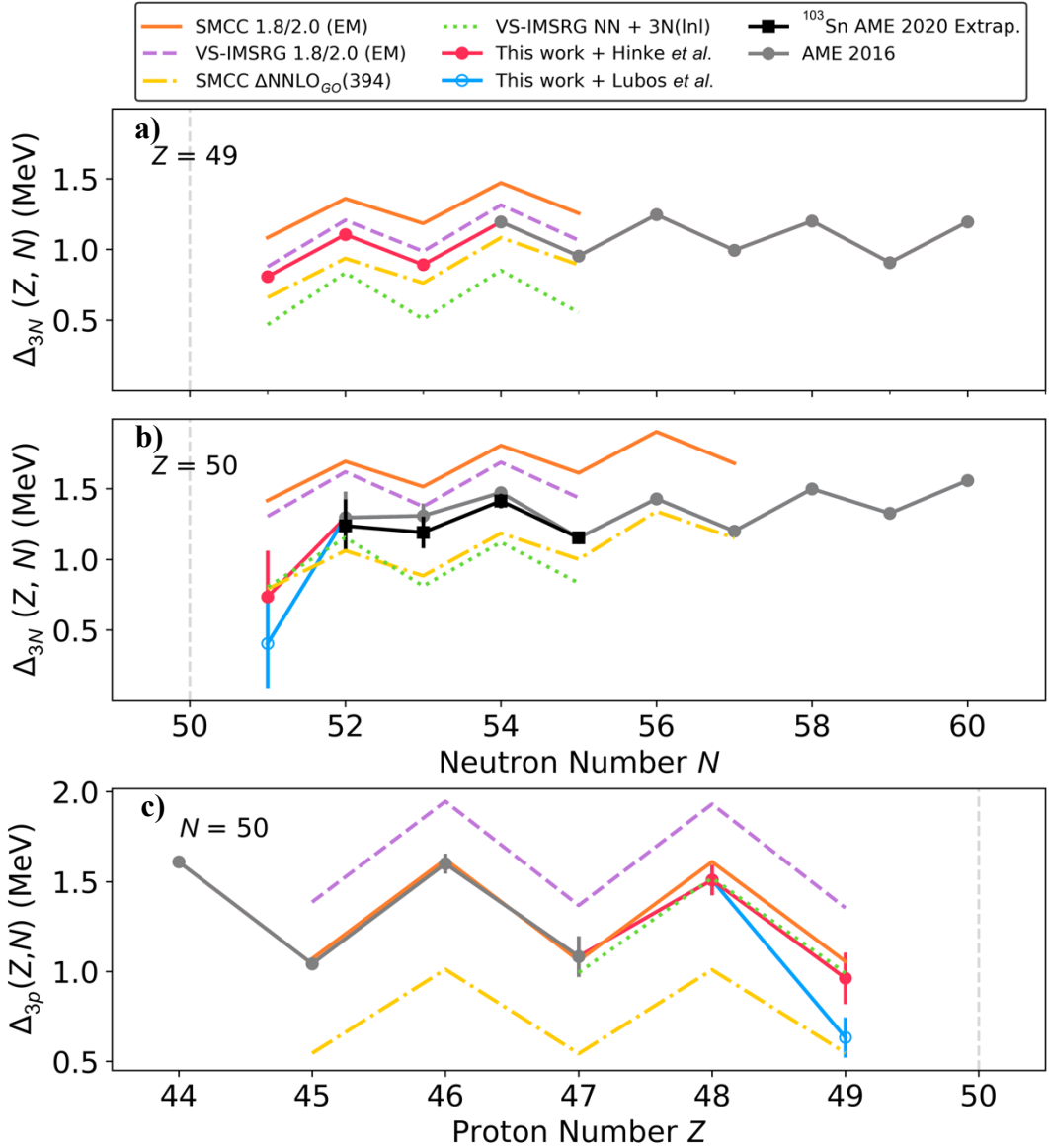

**Figure 3 | Comparison of experimental three-point estimators of the odd-even staggering with theoretical results. a)** Three-point empirical formula of the neutron odd-even staggering in the indium ($Z$ = 49) isotopic chain as a function of the neutron number. **b)** Three-point empirical formula of the neutron odd-even staggering in the tin ($Z$ = 50) isotopic chain as a function of the neutron number. The solid black line represents the same quantity computed considering the extrapolated $^{103}$Sn mass value given in the AME2020[36,37]. **c)** Three-point empirical formula of the proton odd-even staggering along the $N$ = 50 isotonic chain as a function of the proton number. The points resulting from the $^{100}$Sn mass deduced with the $Q$-values from Hinke *et al.*[4] and Lubos *et al.*[5] are plotted using the same colors as in Fig. 2. *(see text for details)*.

Figure 3a presents the experimental three-point empirical formula of the OES, $\Delta_{3n}(Z,N) = 0.5 \times (-1)^N [M_E(Z,N-1) - 2M_E(Z,N) + M_E(Z,N+1)]$ for the odd-$Z$ indium isotopic chain. Figure 3a also shows the trends of $\Delta_{3n}(Z,N)$ calculated with the *ab initio* methods described above. Both many-body methods using the 1.8/2.0 (EM) interaction yield $\Delta_{3n}(Z,N)$ trends that agree with our experimental results. The differences between both methods are within estimated theoretical uncertainties (See Methods for details). Calculations performed with the $\Delta$NNLO$_{GO}$(394) and NN + 3N(lnl) interactions slightly underestimate the energy but closely follow the experimental trend, like the more explored 1.8/2.0 (EM) interaction. All in all, the predictions vary with the choice of many-body method and nuclear Hamiltonian in a range of 500keV but with all methods yielding excellent trends.

Figure 3b shows the experimental trend of $\Delta_{3n}(Z,N)$ for the tin chain (solid grey line). The experimental $N$=53 point in Fig. 3b deviates from the regular odd-even behavior of the three-point empirical formula of the OES. This deviation is most likely explained by the AME2016[21] $^{103}$Sn mass which is known indirectly via its $\beta$-decay link to $^{103}$In[38,39]. In fact, in the latest version of the AME[36,37], this experimental mass value was found to violate the smoothness of the mass surface in this region to such a degree that the evaluators recommended replacing it value by an extrapolated value. The $\Delta_{3n}(Z,N)$ trend for the tin chain obtained with the $^{103}$Sn AME2020 extrapolated value (solid black line in Fig. 3b) appears more regular and is better reproduced by the various theoretical calculations. Hence, as for $Z = 49$, in $Z = 50$ the relative agreement of the theoretical predictions with experiment is good overall. The successful benchmarking of the *ab initio* calculations by the new indium masses give confidence in their predictions towards $^{100}$Sn, only one nucleon away. At $N = 51$, the discrepancy observed between the $Q$-values reported in Hinke *et al.*[4] and Lubos *et al.*[5] is again highlighted, with that of Ref. [4] more in line with our theoretical results. Since the uncertainties of the light tin masses are not as stringent as our new indium results, we also compare our predictions with the three-point proton odd-even staggering as a function of proton number in Figure 3c. Again, our new calculations agree with the experimental trend all the way up to $Z = 48$, yielding a staggering of similar magnitude and differing only in absolute values. At $Z = 49$ the evolution of all theoretical trends clearly favor the Hinke *et al.*[4] $Q$-value over that of Lubos *et al.*[5].

In conclusion, we have measured the most exotic masses accessible with present-day facilities in the $^{100}$Sn region. Our mass measurements of $^{99,100,101g,101m}$In were made possible only by employing the most advanced mass spectrometry techniques. The high-precision data is used to explore the trends of binding energies in the direct vicinity of $^{100}$Sn. The 90-fold improvement in the $^{100}$In mass amplifies a discrepancy existing in the β-decay $Q$-values used to derive the $^{100}$Sn mass. The new precision mass data allows for the first time extending the experimental knowledge of binding energy to only one proton below $^{100}$Sn. The $^{100}$Sn region constitutes the frontier for the fast developing *ab initio* nuclear theoretical approaches. We performed systematic calculations for the first time in the odd-$Z$ indium isotopic chain employing the VS-IMSRG and SMCC methods with three different chiral effective field theory Hamiltonians. Combined with calculations for the neighboring even-$Z$ nuclides, we demonstrate good agreement between experiment and theoretical trends for the three-point neutron and proton odd-even staggering formulations. From these comparisons, we conclude that of the discrepant $Q$-values of Hinke *et al.*[4] and Lubos *et al.*[5], the Hinke *et al.* $Q$-value is favored both by simple extrapolations of experimental trends and by the *ab initio* predictions. Our new mass values thus provide a successful benchmark and increased confidence for these theoretical approaches concerning the iconic $^{100}$Sn.

**Acknowledgements**: We thank the ISOLDE technical group and the ISOLDE Collaboration for their support. We acknowledge the support of the Max Planck Society, the French Institut National de Physique Nucléaire et de Physique des Particules (IN2P3), the European Research Council (ERC) through the European Union's Horizon 2020 research and innovation program (grant agreement No. 682841 "ASTRUm" and 654002 "ENSAR2"), and the Bundesministerium für Bildung und Forschung (grants Nos. 05P15ODCIA, 05P15HGCIA, 05P18HGCIA, and 05P18RDFN1). J.K. acknowledges the support of a Wolfgang Gentner Ph.D. scholarship of the BMBF (05E12CHA). This work was supported by the U.S. Department of Energy, Office of Science, Office of Nuclear Physics, under Awards No. DE-FG02-96ER40963. This material is based upon work supported by the U.S.

Department of Energy, Office of Science, Office of Advanced Scientific Computing Research and Office of Nuclear Physics, Scientific Discovery through Advanced Computing (SciDAC) program under Award Number DE-SC0018223. TRIUMF receives funding via a contribution through the National Research Council of Canada, with additional support from NSERC. Computer time was provided by the Innovative and Novel Computational Impact on Theory and Experiment (INCITE) Program. This research used resources of the Oak Ridge Leadership Computing Facility located at Oak Ridge National Laboratory, which is supported by the Office of Science of the Department of Energy under Contract No. DE-AC05-00OR22725. The VS-IMSRG computations were performed with an allocation of computing resources on Cedar at WestGrid and Compute Canada, and on the Oak Cluster at TRIUMF managed by the University of British Columbia department of Advanced Research Computing (ARC)

**Author Contribution:** MM, DA, JK, PA, IK, YL, VM, TS, AW, and FW performed the experiment. MM, DA, JK and RNW performed the data analysis. KC and SGW setup the resonant laser ionization scheme. WJH performed the update of the AME with the latest experimental results. GH, JDH, GRJ, TM, TP, SRS and ZHS performed the theoretical calculations. KB, VM, DL, AS, LS, KZ and MM prepared the manuscript. All authors discussed the results and contributed to the manuscript at all stages.

**Methods:**

*MR-ToF MS mass measurement and analysis*
The relation between the time of flight t of a singly charged ion of interest and its mass $m_{ion}$ is given by: $t = a\,(m_{ion})^{1/2} + b$ where *a* and *b* are device-specific calibration parameters. These can be determined from the measured flight times $t_{1,2}$ of two reference ions with well-known masses $m_{ion,1}$ and $m_{ion,2}$. From the time-of-flight information of all the singly charged species, the mass of an ion is then calculated from the relation $m_{ion}^{1/2} = C_{ToF}\,\Delta_{ref} + 0.5\,\Sigma_{ref}$ with $\Delta_{ref} = m_{ion,1}^{1/2} - m_{ion,2}^{1/2}$, $\Sigma_{ref} = m_{ion,1}^{1/2} + m_{ion,2}^{1/2}$ and $C_{ToF} = [2t - t_1 - t_2]/[2(t_1 - t_2)]$[24]. The ions' flight times were recorded with a 100 ps resolution. The peaks corresponding to the indium ions of interest were unambiguously identified by their disappearance when blocking the RILIS lasers. The mean of the ToF distribution corresponding to each ion species was estimated using the unbinned maximum-likelihood method, assuming a Gaussian probability density function (PDF). To cope with the pronounced asymmetries observed in the shape of the time-of-flight distribution, a restricted fit range was used (see Fig. 2). The dependence of the time-of-flight fit to these tails was compared to an analysis using the asymmetric PDF from Ref.[40]. The difference between the extracted mean ToF was subsequently treated as a systematic time-of-flight uncertainty and was found to be the dominating contribution in the final uncertainty. When too many ions are trapped in the MR-ToF MS, space-charge effects can cause the time-of-flight difference between two species to shift, affecting the accuracy of the mass determination. To mitigate this effect the count rate was always kept below 8 ions/cycle, which has proven to be a safe limit from previous tests. Nonetheless, count-rate effects were investigated and were found not to be statistically relevant. In the case of [99]In, an additional source of systematic uncertainty was considered. The sensitivity of the extracted time-of-flight to the presence of a possible isomeric state was studied employing a Monte Carlo approach. We assumed that the ratio of ground and isomeric states for [99]In was similar to that observed for [101]In (i.e., 25:1), because the two states in [99]In are expected to have the same spin and parity. Our procedure yields a conservative estimate since the target release efficiencies (expected to be lower for [99]In than [101]In due to shorter half-lives) are not taken into account. The result of this study was treated as an additional systematic uncertainty which was added in quadrature. Note that our MR-ToF MS mass value for [100]In is in good agreement with our Penning-trap value (see Table 1).

*Principle of Penning-trap mass spectrometry*
Penning-trap mass spectrometry relies upon the determination of the free cyclotron frequency

$\nu_C = qB/(2\pi m_{ion})$ of an ion species stored in magnetic field B and charge q. Comparing $\nu_C$ to the frequency $\nu_{C,ref}$ of a species of well-known mass yields the frequency ratio $r = \nu_{C,ref}/\nu_C$ from which the atomic mass value of the ion of interest can be directly calculated. For singly charged ions, the atomic mass of the species of interest is thus expressed as $m_{atom} = r(m_{atom,ref} - m_e) + m_e$, where $m_e$ is the electron mass[41]. As contributions from electron binding energies are orders of magnitude smaller than the statistical uncertainty, they are neglected here.

*ToF-ICR mass measurements and analysis*
The mass of $^{100}$In was measured using the well-established ToF-ICR technique using both the one-pulse excitation[42] and the two-pulse, Ramsey-type excitation[43]. In this method, the free-cyclotron frequency of an ion is directly determined. From one experimental cycle to the next, the frequency of an excitation pulse is varied. Following this excitation, the ions are ejected from trap and their time-of-flight to a downstream micro-channel plate detector is measured. The response of the ions to the applied excitation is a resonant process whose resonance frequency is $\nu_C$ and for which a minimum of the time-of-flight is observed. In the Ramsey scheme, two excitation pulses coherent in phase and separated by a waiting time are applied. The measured Ramsey-type ToF-ICR resonance for $^{100}$In is shown in Fig. 4a. For the same total excitation time, this method offers a three-fold precision improvement when compared to the single-pulse ToF-ICR method. In both cases, the analysis was performed using the EVA analysis software and the various sources of systematic uncertainties were treated according to Ref. [44]. A mass value for $^{101}$In was likewise measured and agrees with a value determined by PI-ICR (see below) within one combined standard deviation.

*PI-ICR mass measurements and analysis*
In order to separate the $A = 101$ isomers, the recently introduced Phase-Imaging Ion-Cyclotron-Resonance technique was used[18]. With this method, the radial frequency of ions prepared on a pure cyclotron or magnetron orbit is determined through the measurement of the phase they accumulate in a time $t_{acc}$ using the projection of their motion onto a position-sensitive multi-channel plate detector. The PI-ICR technique offers several advantages over the regular ToF-ICR technique. Firstly, it is a non-scanning technique which greatly reduces the number of ions required to perform a measurement, i.e., only 5-10 ions are required where a minimum of 50-100 are required for ToF-ICR. While the resolving power of the ToF-ICR method is entirely limited by the excitation time, the resolving power of PI-ICR depends on the observation time and the ion-distribution spot size projected on the detector.

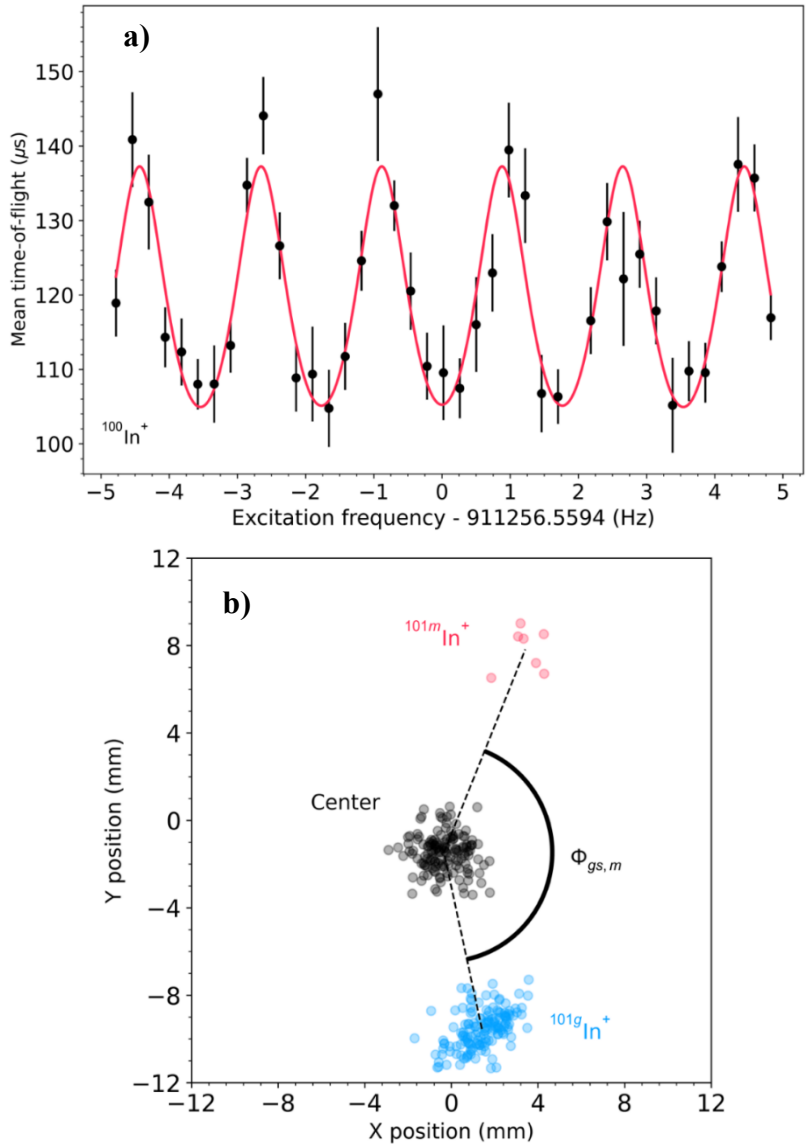

**Figure 4| Overview of experimental results (continued). a)**, Ramsey ToF-ICR resonance of $^{100}$In$^+$ containing about 160 ions. A Ramsey pattern of $T_{RF}^{on}$-$T_{RF}^{off}$-$T_{RF}^{on}$ = 50 ms – 500 ms – 50 ms was used for this measurement. The solid red line corresponds to the least-square adjustment of the theoretical line shape to the data. **b)**, PI-ICR ion-projection image of $^{101}$In$^+$. In a phase-accumulation of about 65 ms a mass resolving power in excess of $10^6$ was reached allowing for the ground (blue) and isomeric (red) states to be separated by the angle $\Phi_{gs,m}$ which directly determines the nuclear excitation energy. The centre (black) of the projected ion motion is obtained in a separate measurement.

A three-step measurement scheme allows for the direct determination of $\nu_c$. First, a position measurement is performed without preparing the ions on a specific motion radius, yielding the position of the center of the ions' motion. In a second step, the ions are prepared on a pure magnetron orbit, left to evolve freely during a time $t_{acc}$ and their position is measured. Finally, the ions are prepared on a pure reduced cyclotron orbit, left to evolve freely during the same time $t_{acc}$ and their position is again measured. The integer number of revolutions $n_-$ and $n_+$, performed in steps 2 and 3 respectively, the phase accumulation time $t_{acc}$ and the angle $\Phi$ between the ions' positions obtained in steps 2 and 3 can be related to $\nu_c$ following the relation $\nu_c = [2\pi(n_- + n_+) + \Phi]/ t_{acc}$. In step 3, the phase accumulation is performed at the modified cyclotron frequency, so is mass dependent. The position of each ion spot was extracted using the unbinned maximum-likelihood method, assuming a 2D multivariate Gaussian distribution[45]. Figure 4 shows a typical PI-ICR image obtained in step 3 after ~65 ms of phase accumulation. If in principle the angle $\Phi_{gs,m}$ between the ground and isomeric state directly reflects the energy difference between the two states, the mass of both states was measured separately in order to mitigate systematical effects. The PI-ICR method was used to study the isomeric composition of the $^{100}$In beam. Hence, we can exclude the presence of a long-lived state with an excitation energy larger than 20 keV in the $^{100}$In beam delivered to ISOLTRAP's measurement Penning trap.

*VS-IMSRG calculations*
The VS-IMSRG calculations[26,46] were performed in a spherical harmonic-oscillator basis including up to 15 major shells in the single-particle basis with an oscillator frequency $\hbar\omega$ = 16 MeV. The 3N interaction configurations were restricted up to $e_1+e_2+e_3 \leq E_{3max}$ = 16 for the 1.8/2.0(EM) interaction (to compare with SMCC calculations) and $E_{3max}$ = 22 for the NN + 3N(lnl) interaction. We first transform to the Hartree-Fock basis, then use the Magnus Formulation of the IMSRG[47] to construct an approximate unitary transformation to decouple a $^{78}$Ni core with a proton $p_{1/2}$, $p_{3/2}$, $f_{5/2}$, $g_{9/2}$ and neutron $s_{1/2}$, $d_{3/2}$, $d_{5/2}$, $g_{7/2}$, $h_{11/2}$ valence space. Using the ensemble normal-ordering introduced in[26], we approximately include effects of 3N interactions between valence nucleons, such that a specific valence-space Hamiltonian is constructed for each nucleus to be studied. The final diagonalization is performed using the KSHELL shell-model code[48]. To estimate theoretical uncertainties in this framework, we note that in the limit of no IMSRG truncations, results would be independent of the chosen reference state for the ensemble normal ordering procedure. Therefore, we examine the reference-state dependence of the observables discussed above. Normal ordering with respect to either a filled neutron $g_{7/2}$ or $d_{5/2}$ orbits, we find approximately 1MeV uncertainty for absolute or one-neutron separation energies. However, for all quantities shown in Fig. 3, this estimated uncertainty is approximately 0.1 MeV.

*SMCC calculations*
The SMCC approach generates effective interactions and operators through the decoupling of a core from a valence space. We start from a single Hartree-Fock $^{100}$Sn reference state, computed in a harmonic oscillator basis comprising up to 11 major oscillator shells and an oscillator frequency $\hbar\omega$ = 16 MeV. The 3N interaction was restricted to $E_{3max} = 16\hbar\omega$. The doubly closed-shell $^{100}$Sn core is decoupled by a coupled-cluster calculations including singles, doubles and the leading-order triples excitations (CCSDT-1 approximation). We note that triples excitations were performed in the full model-space, without any truncations. Thisx was made possible by employing the Nuclear Tensor Contraction Library (NTCL)[49] developed to run at scale on Summit, the U.S. Department of Energy's 200 petaflop supercomputer operated by the Oak Ridge Leadership Computing Facility (OLCF) at Oak Ridge National Laboratory. The SMCC calculations then proceed via a second similarity transformation that decouples a particle-hole valence-space defined by the proton $pfg_{9/2}$ holes and neutron $g_{7/2}sd$ single-particle states. The SMCC decoupling only includes the one- and two-body parts of the CCSDT-1 similarity-transformed Hamiltonian. To estimate theoretical uncertainties, we note that the calculation of doubly-magic nuclei such as $^{100}$Sn or $^{78}$Ni and their neighbors are ideally suited for the coupled-cluster method, because the reference state is closed shell[2,46]. Comparison of the SMCC results for $^{101}$Sn with those from Ref[2] exhibit differences on single-particle energies of about 0.2 MeV. We therefore estimate that our theoretical uncertainties on $\Delta_{3n}(Z,N)$ are about ±0.2 MeV.

Code Availibity:

The analysis codes used for the ToF-ICR and MRToF-MS data are available from the corresponding author upon reasonable request. A second MRToF-MS analysis code used in this study is available at this address: https://github.com/jonas-ka/mr-tof-analysis. The PI-ICR analysis code used in this study is available at this address: https://github.com/jonas-ka/pi-icr-analysis. The code used for the VS-IMSRG calculations is available at https://github.com/ragnarstroberg/imsrg. The source code of KSHELL is available in Ref.[48].